\documentclass{pasj01}
\Received{$\langle$reception date$\rangle$}
\Accepted{$\langle$acception date$\rangle$}
\Published{$\langle$publication date$\rangle$}

\begin{document}

\title{The Formation of Massive Molecular Filaments and Massive Stars Triggered by a MHD Shock Wave}
\author{Tsuyoshi Inoue\altaffilmark{1,2}, Patrick Hennebelle\altaffilmark{3}, Yasuo Fukui\altaffilmark{1}, Tomoaki Matsumoto\altaffilmark{4}, Kazunari Iwasaki\altaffilmark{5} and Shu-ichiro Inutsuka\altaffilmark{1}}%
\altaffiltext{1}{Department of Physics, Graduate School of Science, Nagoya University, Furo-cho, Chikusa-ku, Nagoya 464-8602, Japan}
\altaffiltext{2}{Division of Theoretical Astronomy, National Astronomical Observatory of Japan, Osawa 2-21-1, Mitaka, Tokyo 181-0015, Japan}
\altaffiltext{3}{Laboratoire AIM, Paris-Saclay, CEA/IRFU/SAp -- CNRS -- Universit\'e Paris Diderot, 91191 Gif-sur-Yvette Cedex, France}
\altaffiltext{4}{Faculty of Sustainability Studies, Hosei University, Fujimi, Chiyoda-ku, Tokyo 102-8160, Japan}
\altaffiltext{5}{Department of Earth and Space Science, Osaka University, Toyonaka, Osaka 560-0043, Japan}
\email{tsuyoshi.inoue@nagoya-u.jp}


\maketitle

\begin{abstract}
Recent observations suggest that intensive molecular cloud collision can trigger massive star/cluster formation.
The most important physical process caused by the collision is a shock compression.
In this paper, the influence of a shock wave on the evolution of a molecular cloud is studied numerically by using isothermal magnetohydrodynamics (MHD) simulations with the effect of self-gravity.
Adaptive-mesh-refinement and sink particle techniques are used to follow long-time evolution of the shocked cloud.
We find that the shock compression of turbulent inhomogeneous molecular cloud creates massive filaments, which lie perpendicularly to the background magnetic field as we have pointed out in a previous paper.
The massive filament shows global collapse along the filament, which feeds a sink particle located at the collapse center.
We observe high accretion rate $\dot{M}_{\rm acc}> 10^{-4}$ M$_{\rm sun}$ yr$^{-1}$ that is high enough to allow the formation of even O-type stars.
The most massive sink particle achieves $M>50$ M$_{\rm sun}$ in a few times $10^5$ yr after the onset of the filament collapse.
\end{abstract}

\section{Introduction}
How massive stars are formed is a long-standing issue of astrophysics (e.g., Zinnecker \& Yorke 2007; Tan et al.~2014).
Theoretical studies have shown that massive stars can be formed if we set up a massive gravitationally bound core or clump that contains many Jeans masses in a sub-parsec region (Nakano et al.~2000; McKee \& Tan 2002; Yorke \& Sonnhalter 2002; Krumholz et al.~2009; Bonnell et al.~2001; Kuiper et al.~2010), although it depends on the level of turbulence, magnetization, and density structure of the initial core/clump (Peretto et al.~2007; Commercon et al.~2011; Myers et al.~2013; Peters et al.~2011; Hennebelle et al.~2011; Girichidis et al.~2011).
Recent observations for environmental conditions around massive star/cluster emphasize the importance of molecular cloud collision as a triggering mechanism of the massive star formation (Hasegawa et al.~1994; Furukawa et al.~2009; Ohama et al.~2010; Torii et al.~2011, 2015; Fukui et al.~2014, 2015, 2016; Dobashi et al.~2014; Higuchi et al.~2014; Tsuboi et al.~2015).
Many new observational evidences of the collision triggered massive star formation are reported in this special issue (Fukui et al. 2017a, b, c; Hayashi et al. 2017; Nishimura et al. 2017a, b; Ohama et al. 2017a, b; Sano et al. 2017; Torii et al. 2017; Tsutsumi et al. 2017; Kohno et al. 2017).
These works revealed that the cloud collisions that caused massive star/cluster formation were very intensive with relative velocity on the order of 10 km s$^{-1}$, which inevitably cause strong shock compression.

In theoretical side, molecular cloud collision have been studied mostly by using numerical simulations.
Pioneering work by Habe \& Ohta (1992) showed that the cloud collision forms a dense shock compressed layer in which star formation is triggered even if initial clouds are gravitationally stable (see also, Anathpindika 2010; Takahira et al.~2014; Matsumoto et al.~2015; Balfour et al.~2017; Shima et al.~2017).
The above theoretical studies are based on hydrodynamics simulations omitting the effects of magnetic field.
In molecular cloud, because radiative cooling is efficient, we often treat a cloud as an isothermal gas that leads shock compression ratio $r\simeq M_{\rm s}^2$ in the case of no magnetic field, where $M_{\rm s}$ is the sonic Mach number.
Given that the typical observed collision velocity $\sim 10$ km s$^{-1}$ and sound speed 0.2 km s$^{-1}$ ($M_{\rm s}\sim 50$), the density of shocked layer can be $n=n_0\,M_{\rm s}^2\gtrsim 10^6$ cm$^{-3}$, if we take the initial cloud density $n_0=10^3$ cm$^{-3}$.
Since the thermal Jeans mass is a decreasing function of the density, the shocked gas created by the cloud collision would be a favorable site of low-mass star formation, which is consistent with the results of the above mentioned simulations.

The effect of magnetic field can drastically change the physical state of the shocked cloud.
The density of the post shock gas is not drastically enhanced as above, because magnetic pressure prevents over-contraction behind the shock.
The shock compression ratio can be expressed by $r\simeq \sqrt{2}\,M_{\rm A}$ where $M_{\rm A}=\sqrt{4\,\pi\,\rho_0}\,v_{\rm 0}/B_{t,0}$ is the Alfv\'en Mach number composed of the upstream magnetic field component perpendicular to the shock normal $B_{t,0}$.
This compression ratio is applicable when the Alfv\'en Mach number is larger than the sonic Mach number: $M_{\rm A}>M_{\rm s}$.
By rewriting this inequality, we obtain the range of the magnetic field strength in which we should apply the MHD formula of the compression ratio:
\begin{eqnarray}
B_{t,0}^{\rm cr}&>&\sqrt{8\,\pi\,\rho_{0}}\frac{c_{\rm s}^2}{v_{\rm sh}}\nonumber\\
&\simeq& 0.1\,\,\mu\mbox{G}\,\left( \frac{v_{\rm sh}}{10\mbox{ km s}^{-1}} \right)^{-1}
\left( \frac{n_{0}}{10^3\mbox{ cm}^{-1}} \right)^{1/2}\left( \frac{c_{\rm s}}{0.2\mbox{ km s}^{-1}} \right)^2.
\end{eqnarray}
This indicates that we have to consider the effect of magnetic field quite generally in molecular clouds.

Based on MHD simulations, Chen \& Ostriker (2014, 2015) reported that typical mass of gravitationally bound cores formed by the cloud collision is determined by the condition of the mass-to-magnetic-flux-ratio behind the shock equals unity.
Their result indicates that typical mass of the stars formed by the cloud collision remains small.
Wu et al. (2017a, 2017b) showed that the cloud collision enhances overall star formation activity even if the effect of magnetic field is taken into account.
Inoue \& Fukui (2013) pointed out that, although most created cores are low-mass, the most massive core formed by the cloud collision can be as large as 200 M$_{\rm sun}$ which is massive enough to evolve into massive star(s).
They also suggested that, because of enhanced magneto-sonic speed (or Alfv\'en velocity $c_{\rm A}$) and turbulence ($\Delta v$) behind the shock, the effective Jeans mass ($\sim G^{-3/2}\,\rho^{-1/2}\,\{c_{\rm s}^2+c_{\rm A}^2+\Delta v^2 \}^{3/2}$) and the mass accretion rate ($\sim G^{-1}\{c_{\rm s}^2+c_{\rm A}^2+\Delta v^2 \}^{3/2}$) can take a much larger value in the shocked region created by the cloud collision.
In addition to this, they found that the massive cores created by the cloud collision are generally embedded in massive filaments (see also, Vaidya et al.~2013; Matsumoto et al.~2015), which resemble the results of recent observations of massive core formation sites (Galv\'{a}n-Madrid et al.~2010; Peretto et al.~2013, 2014; Fukui et al.~2015).
 
However, Inoue \& Fukui (2013) didn't follow the collapse phase of the massive filament/core because of limited resolution.
In order to clarify whether the cloud collision can induce massive star formation, we need to perform a seamless simulation from the cloud collision phase to the massive core collapse phase via the massive filament formation.
For this purpose, in this paper, we perform MHD simulations of a cloud collision with adaptive-mesh-refinement (AMR) and sink particle techniques by employing the SFUMATO code developed by Matsumoto (2007) and Matsumoto et al. (2015).

 \begin{figure}[t]
\includegraphics[scale=0.5]{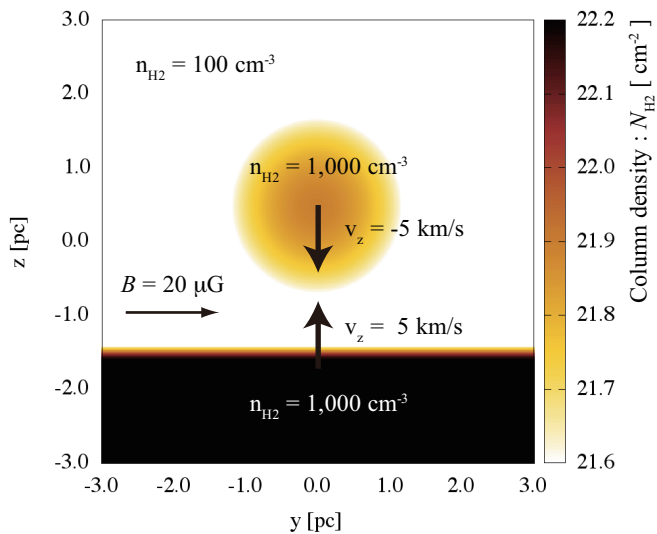}
\caption{\label{f1}
Schematic illustration of the initial setting.
A small cloud of a total mass $M = 838\,\mbox{M}_{\rm sun}$ collides with a larger cloud that is expressed as a plane parallel sea of dense gas.
Turbulent velocity field with dispersion $\Delta v=1.5$ km s$^{-1}$ is set in the small cloud.
From our initial setting, we can study the cloud collision of the case that one cloud is much larger than the other one.
We can also learn the case when the cloud is crushed by plane parallel shock.
}\end{figure}

\section{Numerical Setup}
We solve the isothermal MHD equations with self-gravity by using the SFUMATO code (Matsumoto 2007), in which 
a block-structured AMR technique is used and Poisson’s equation is solved by the multi-grid method.
The MHD equations are integrated by a Godunov type scheme using approximate HLLD Riemann solver (Miyoshi \& Kusano 2005) with the third-order accuracy in space and the second order in time.
The divergence free condition is ensured by using a divergence cleaning method developed by Dedner et al.~(2002).

As an initial cloud, we set a gas sphere with density $n=\rho/m=10^3$ cm$^{-3}$, isothermal sound speed $c_{\rm s}=0.3$ km s$^{-1}$ and radius $r=1.5$ pc, where $m=2.4\,m_{\rm proton}$ is used as a mean mass of molecular gas particles and total mass of the sphere is $M = 838\,\mbox{M}_{\rm sun}$.
Following Larson's law (Larson 1981), we set a turbulent velocity field of the velocity dispersion $\Delta v=1.5$ km s$^{-1}$ with the velocity power spectrum $v_k^2\propto k^{-4}$ (equivalent to the scale-velocity dispersion relation of $\Delta v_l\propto l^{1/2}$).
 
 The numerical domain is set to be $L_{\rm box}=6.0$ pc cubic where the coordinate origin $(0,0,0)$ is set at its center with base resolution $\Delta x_0=L_{\rm box}/512$.
We set the center of the turbulent sphere at $\vec{r}_{\rm sc}=(0,0,0.5\,\mbox{pc})$ and set dense gas of $n=10^3$ cm$^{-3}$ in the region $z<z_{\rm lc}= -1.5$ pc.
In the rest of the box, the less dense gas of $n=10^2$ cm$^{-3}$ is filled as an atmospheric medium.
To induce cloud collision, we set a converging flow as $v_{z}=-5.0\,\mbox{km s}^{-1}\,\tanh(z-z_{\rm lc})$.
The periodic boundary condition is used for $x,\,y=\pm 3.0$ pc boundary planes, and free boundary condition is set for $z=\pm 3.0$ pc boundary planes.
We initially impose uniform magnetic field of $\vec{B}=(0,20.0\,\mu\mbox{G},0)$, which is perpendicular to the collision direction.
This magnetic field strength is consistent with observed strengths in molecular clouds (Crutcher et al.~2010).
Under this magnetic field, the mass-to-magnetic-flux-ratio of the small cloud is about 2.5 times the critical ratio, indicating the initial cloud is magnetically supercritical (Mouschovias \& Spitzer 1976).
The schematic illustration of the initial setting is shown in Figure \ref{f1}.
Since the magnetic field component perpendicular to the collision direction is expected to be highly amplified by the shock compression (while the parallel component is not), the initial $z$-component of the magnetic field would play minor role even if it is given.
 
We use the Jeans criterion (Truelove 1997) as the AMR condition: $\Delta x\le \lambda_{\rm J}/f$, where $\lambda_{\rm J}=\pi^{1/2}\,c_{\rm s}/(G\,\rho)^{1/2}$ is the Jeans length and $f=8$ is used for the fiducial run.
The refinement is allowed until the finest resolution reaches $\Delta x_{\rm min}=L_{\rm box}/4096\simeq 1.5\times 10^{-3}$ pc, where the local density reaches $n_{\rm cr}\simeq 8.1\times 10^6$ cm$^{-3}$ in such resion for the fiducial run.
When the local density increases more than $n_{\rm cr}$, we introduce a sink particle if such a region satisfies four more criteria: (i) the local potential minimum criterion, (ii) the negative velocity divergence criterion, (iii) the negative eigenvalue criterion of symmetric parts of velocity gradient tensor, and (iv) the negative total energy criterion within the sink radius of $4\,\Delta x_{\rm min}$ (Federrath et al.~2010; Matsumoto et al.~2015).

With our initial setting, we can study the cloud collision of the case that one cloud is much larger than the other one.
We can also learn the case when the cloud is crushed by a plane parallel shock wave.
Note that we do not set the initial turbulence in the larger cloud for $z<z_{\rm lc}$ in order to focus only on the shock crushed small cloud, which evolves into dense filaments as pointed out by Inoue \& Fukui (2013).
If we take into account the initial turbulence in the larger cloud, we would get additional star formation in shocked larger cloud region.

 \begin{figure}[t]
\includegraphics[scale=0.51]{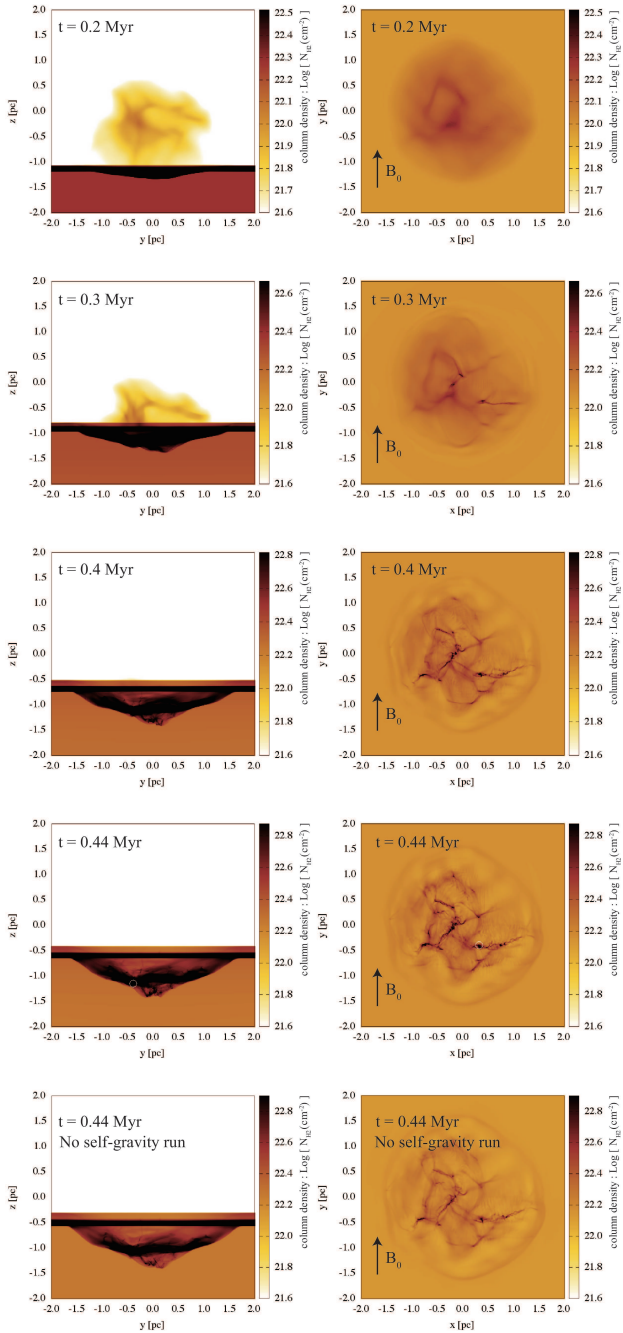}
\caption{\label{f2}
Snapshots of column density structure at $t=0.2,\,0.3,\,0.4,$ and $0.44$ Myr.
Left panels are the column density along the $x$-axis and right panels are those along the $z$-axis.
In the last row, we also plot the result of the run without self-gravity.
At $t=0.44$ Myr the first sink particle is just created at $(x,\,y,\,z)=(0.307,\,-0.405,\,-1.13)$.
}\end{figure}

 \begin{figure}[t]
\includegraphics[scale=0.51]{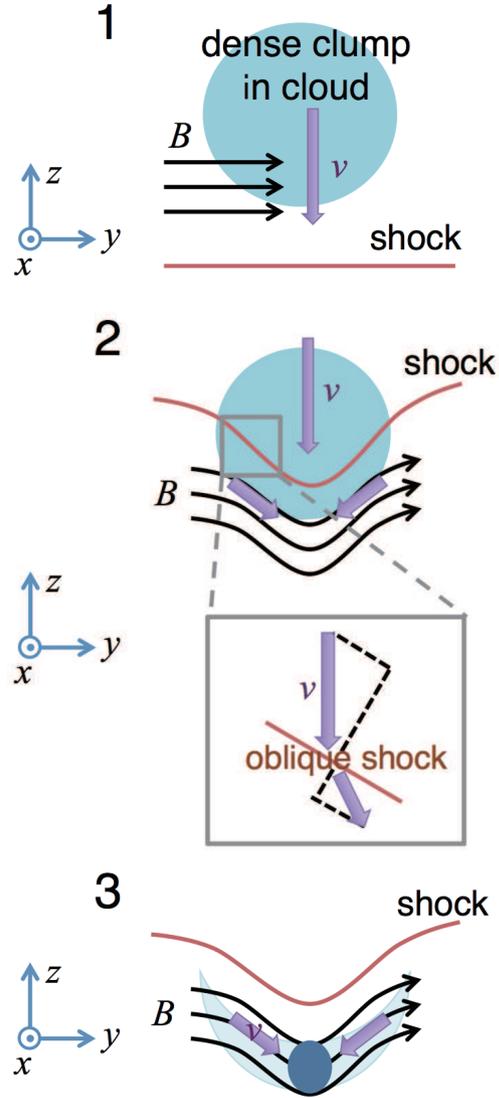}
\caption{\label{f3}
Illustration of the physical mechanism of the filament formation proposed by Inoue \& Fukui (2013), where the evolution of a dense clump after shock compression is considered.
A filament perpendicular to the plane of the paper is created at the stage 3.
Here the dense clump under consideration is created by the initial turbulence before the collision.
Thus, in the simulation, many dense clumps exist before the shock sweeps the small cloud, and the existence of many clumps leads to the formation of many filaments in the shock crushed cloud.
}\end{figure}

 \begin{figure}[t]
\includegraphics[scale=0.4]{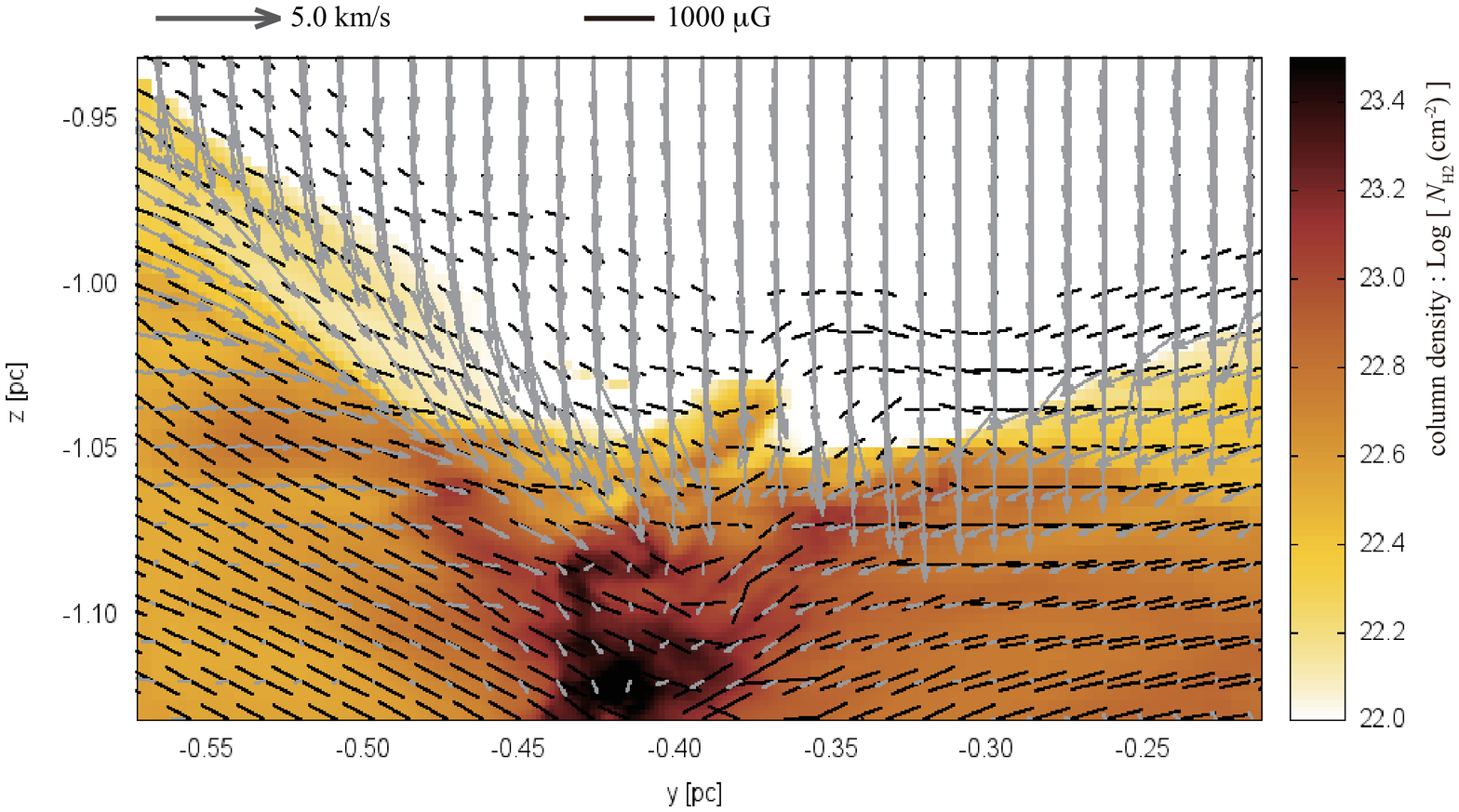}
\caption{\label{f4}
Local structure of a filament formation site at $t=0.44$ Myr.
The color represents the column density and the black lines show mass-weighted average magnetic field projected on to the plane.
The mass-weighted average velocity field is shown as gray vectors.
Note that the formed filament orients perpendicular to the paper.
}\end{figure}

\section{Results}
\subsection{Filament Formation Phase}\label{fil}
After the collision, the turbulent inhomogeneous cloud is compressed by the shock wave and filamentary structures are quickly developed in the crushed cloud.
In Figure \ref{f2} we show snapshots of the column density structure at $t=0.2,\,0.3,\,0.4,$ and $0.44$ Myr.
Left panels are the column density along the $x$-axis and right panels are those along the $z$-axis.
In the last row, we also plot the result of the run without self-gravity.
At $t=0.44$ Myr the first sink particle is created at $(x,\,y,\,z)=(0.307,\,-0.405,\,-1.13)$, indicating that the effect of self-gravity have just started to play role in the densest region at this time, and in most regions, gas motion is still determined as a MHD phenomenon.
In fact, by comparing the results with and without the effect of self-gravity in Figure \ref{f2}, we can confirm that the filamentary structures are not a consequence of the self-gravity.

Inoue \& Fukui (2013) pointed out that filamentary structures are generated when an inhomogeneous cloud is swept by a MHD shock wave (see also, Vaidya et al.~2013; Inutsuka et al.~2015).
In the following of this section, we show that the filament formation mechanism in the present simulation is the same as that proposed by Inoue \& Fukui (2013).
Figure \ref{f3} is the illustration about the physical mechanism of the filament formation by Inoue \& Fukui (2013), where the evolution of a dense clump after shock compression is considered as follows:
When the shock wave induced by the collision hits a dense clump in turbulent cloud, the shock front is deformed, because the shock speed is decelerated in the dense region.
At the deformed shock wave, gas flow is kinked as depicted in Figure \ref{f3} due to the oblique shock effect.
This is because the velocity (or momentum flux) tangential to the shock is conserved, while the normal velocity is stalled across the shock\footnote{This explanation is exact for hydrodynamic shock, but not exact for MHD shocks. In the MHD, the tangential momentum flux involves the magnetic tension term and thus the tangential velocity is not exactly conserved across the shock. However, so far as the shock wave is a MHD fast shock, the flow kink at the shock front as depicted in Figure \ref{f3}.}.
As a consequence of the flow kink, focusing flows are generated behind the shock which further compress the clump, i.e., the clump is compressed not only by the shock but also by the post shock focusing flows.
Because of the strong magnetic field behind the shock, the focusing flows can be induced only along the post shock magnetic field.
This indicates that the filamentary structures perpendicular to the magnetic field (or perpendicular to the plane of the paper in Figure \ref{f3}) are created.

Note that the dense clump under consideration in Figure \ref{f3} is created by the initial turbulence before the collision.
Thus, many dense clumps exist in the small cloud, and the existence of the clumps leads to the formation of many filaments in the shock crushed cloud.

To confirm the above mechanism, we plot a local structure of a filament formation site at $t=0.44$ Myr in Figure \ref{f4}, where the first sink particle is just created at $(x,\,y,\,z)=(0.307,\,-0.405,\,-1.13)$ in this filament formation site.
The color represents the column density, and the black lines show the mass-weighted average magnetic field projected on to the plane.
The mass-weighted average velocity field is shown as gray vectors.
We can see the fairly similar structure as in the panel 3 of Figure \ref{f3} in actual data.
As we shall discuss in the next section, the first sink particle created in this filament evolve into the most massive one, indicating that the most massive core is created in the filament that is formed by Inoue \& Fukui (2013) mechanism.

 \begin{figure*}[t]
\includegraphics[scale=0.85]{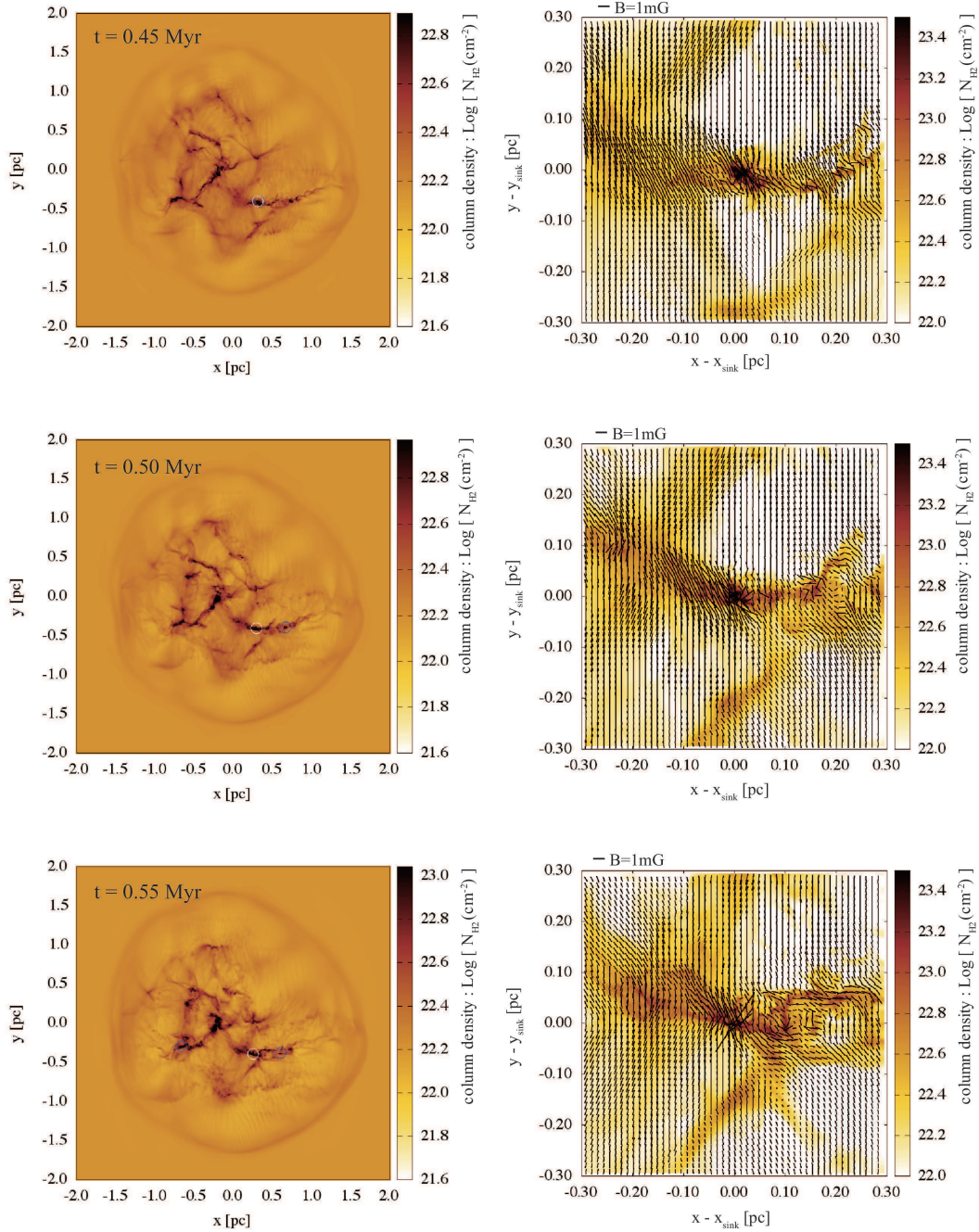}
\caption{\label{f5a}
Snapshots of column density structure at $t=0.45,\,0.50,$ and $0.55$ Myr.
Left panels are the column density along the $z$-axis and right panels are enlarged view around the most massive sink particle.
The position of the most massive sink is indicated as a white open circle in the left panels, and blue circles indicate the positions of less massive sinks.
}\end{figure*}

 \begin{figure*}[t]
\includegraphics[scale=0.85]{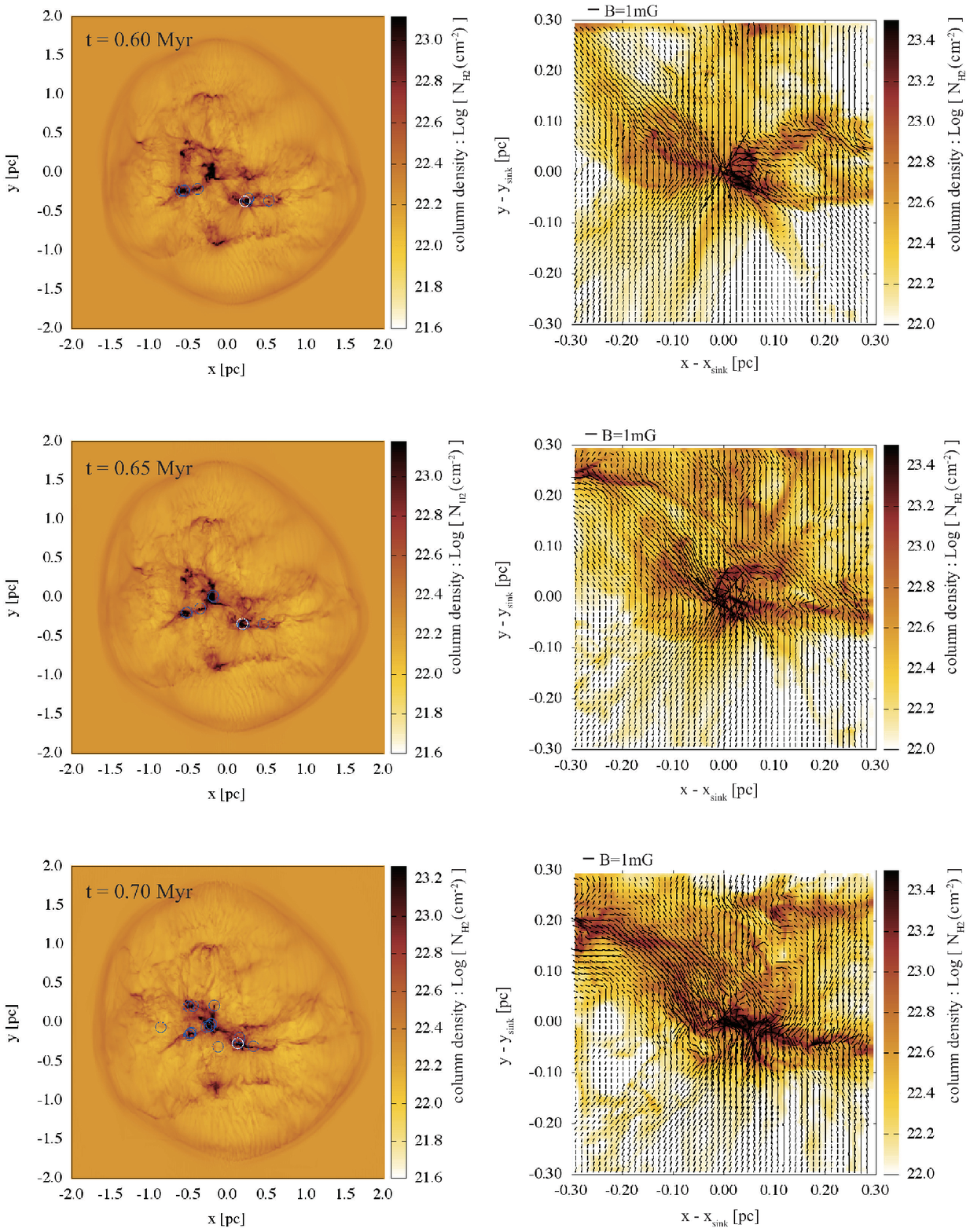}
\caption{\label{f5b}
Same as Figure \ref{f5a}, but for the structures at $t=0.60,\,0.65,$ and $0.70$ Myr.
}\end{figure*}

\subsection{Filament Collapse Phase}
In Figure \ref{f5a} and \ref{f5b} we show snapshots of the column density structure at $t=0.45,\,0.50,\,0.55,\,0.60,\,0.65$ and $0.70$ Myr.
The left panels are the column density along the $z$-axis and right panels are enlarged view around the most massive sink particle.
The position of the most massive sink is indicated as a white open circle in the left panels, and blue circles indicate the positions of less massive sinks.
One can see the animation of the corresponding evolution in supplementary movie 1.
We see that the filament that bears the most massive sink shows longitudinal global collapse along its major axis.

The top-right panel of Figure \ref{f5a} shows that the massive filament is running almost along the $x$-direction.
To estimate the line-mass of the massive filament, we have computed the total mass of the dense gas with $n\ge 10^4$ cm$^{-3}$ inside the cylindrical region of $V=\{ (x,y,z)|\,|x-x_{\rm sink}|\le 0.25\mbox{ pc},\, [(y-y_{\rm sink})^2+(z-z_{\rm sink})^2]^{1/2}\le 0.1\mbox{ pc}\}$ at $t=0.45$ Myr.
The resulting mass $M_{V}=40$ M$_{\rm sun}$ indicates that the line-mass can be $\lambda\simeq 80$ M$_{\rm sun}$ pc$^{-1}$.

It is known that an unmagnetized filament whose line-mass is larger than the critical mass 
$\lambda_{\rm max}=2\,c_{s}^{2}/G$ 
cannot have equilibrium structure, and thus is gravitationally unstable (Stodolkiewicz 1963; Ostriker 1964).
Recently, Tomisaka (2014) showed that the maximum line-mass of a filament that is threaded by magnetic field perpendicular to the filament is modified to be
\begin{equation}
\lambda _{\rm max} \simeq 0.24 \Phi_{\rm cl}/G^{1/2}+1.66\,c_{s}^{2}/G,
\end{equation}
where $\Phi_{\rm cl}$ represents one half of the magnetic flux threading the filament per unit length (or one half of the magnetic field strength times the width of the filament).

As we have shown in the Section \ref{fil}, the filaments in the present simulation are formed by the mechanism proposed by Inoue \& Fukui (2013),
in which it is shown that the magnetic field strength of the filaments ($B_{\rm fil}$) can be predicted accurately by the shock compression value of the perpendicular component of the magnetic field to the shock normal:
\begin{eqnarray}\label{Bfil}
B_{\rm fil}&\simeq& B_{1}=r\,B_{0}\nonumber \\
&=&\left[\left\{ 2\,M_{\rm A}^2+(\beta+1)^{2}/4 \right\}^{1/2}-(\beta+1)/2\right]\,B_{0}\nonumber\\
&\simeq& \sqrt{2}\,M_{\rm A}\,B_{0}\nonumber\\
&\simeq& 300\,\mu\mbox{G}\,\left( \frac{n_{0}}{10^3\mbox{ cm}^{-3}} \right)^{1/2}\,\left( \frac{v_{\rm sh}}{10\mbox{ km s}^{-1}} \right),
\end{eqnarray}
where $r$ is the compression ratio, the subscripts 0 (1) represents the preshock (postshock) value in the shock rest frame, $M_{\rm A}\equiv v_{0}/(B_0/\sqrt{4\pi\,\rho_{0}})$ is the Alfv\'enic Mach number, $\beta \equiv 8\,\pi\,c_{\rm s}^2\,\rho_0/B_0^2$ is the upstream plasma beta, and in the third line we have used $M_{\rm A}\gg\beta$.
In the eq.~(\ref{Bfil}), we have employed the shock-jump-condition for the perpendicular, isothermal MHD shock.
The result is not substantially modified, even if there is a non-zero magnetic field component parallel to the shock normal\footnote{From the shock-jump-condition with the non-zero parallel magnetic field, we can express the amplification factor of the perpendicular magnetic field as: $r_{B}=\{ 2\,M_{\rm A}^2\,(M_{\rm s}^2-r)\,(r-1)/(M_{\rm s}^2\,r)+1 \}$.
In the case of the strong isothermal MHD shock, the condition $M_{\rm s}^2\gg r\sim M_{\rm A}>1$ is satisfied in wide range of parameter that leads the same expression to the eq.~(\ref{Bfil}).}.
To estimate the filament magnetization in the simulation, we have computed the average magnetic field strength in the region $V$ given above at $t=0.45$ Myr: $\langle |B| \rangle_{V}=670$ $\mu$G.
This is somewhat stronger than that of the eq.~(\ref{Bfil}), because the magnetic field got additional amplification due to the gravitational contraction around the sink (see, top right panel of Figure \ref{f5a}).

If we use the typical magnetic field strength of the filament before collapse as $B_{\rm fil}\sim 300\,\mu$G and use width of the filament as $w_{\rm fil}\sim 0.1$ pc (Arzoumanian et al.~2011), the maximum line mass is estimated to be
\begin{eqnarray}\label{Mfil}
\lambda_{\rm max}&\simeq& 67\mbox{ M}_{\rm sun}\,\mbox{pc}^{-1}\, (B_{\rm fil}/300 \mu\mbox{G})\,(w_{\rm fil}/0.1\, \mbox{pc})\nonumber\\
&&+35\mbox{ M}_{\rm sun}\,\mbox{pc}^{-1}\,(c_{\rm s}/0.3\mbox{ km s}^{-1})^2,
\end{eqnarray}
where corresponding average column density of the filament is $\simeq 4.3\times10^{22}$ cm$^{-2} (w_{\rm fil}/0.1\mbox{ pc})^{-1}$.
This critical line mass is consistent with the line-mass of the filament in the present simulation, suggesting that we can have much larger critical line-mass than the pure thermal case thanks to the strong magnetic field behind the shock wave.
If such a massive filament collapses globally, we can naturally expect massive star formation.

 \begin{figure}[t]
\includegraphics[scale=0.65]{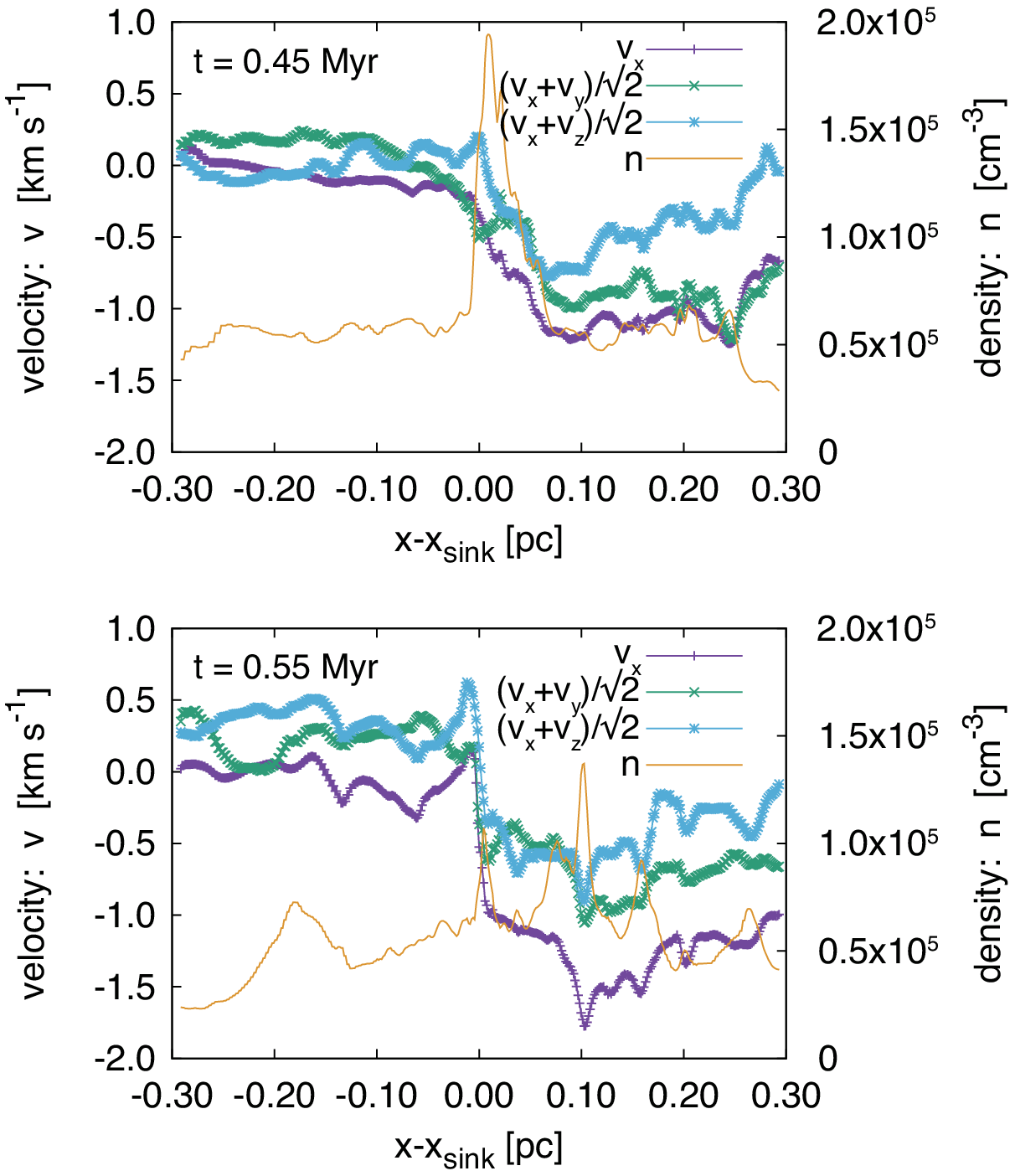}
\caption{\label{f6}
Velocity structure along the $x$-axis (filament) that passes by the sink position.
Top and bottom panels respectively show the results of $t=0.45$ and $0.55$ Myr.
Purple lines show density weighted average $v_x$ around the $x$-axis: $\langle v_x(x) \rangle_{y,z}\equiv\int_{S}\,\rho\,v_x\,dy\,dz/\int_{S}\,\rho\,dy\,dz$ where $S=\{(y,z)|\,[(y-y_{\rm sink})^2+(z-z_{\rm sink})^2]^{1/2}\le0.1\,\mbox{pc}\}$.
The green and blue lines indicate the velocity structures of $\langle v_x(x)+v_y(x) \rangle_{y,z}/\sqrt{2}$ and $\langle v_x(x)+v_z(x) \rangle_{y,z}/\sqrt{2}$, respectively.
If we observe this filament by using some line emissions from an oblique direction with an angle of 45$^\circ$ to the $z$-axis in the $x$-$y$ plane ($x$-$z$ plane), we would find the structure similar to the green (blue) line in position-velocity space.
Yellow lines show average number density: $\langle n(x) \rangle_{y,z}\equiv\int_{S}\,n\,dy\,dz/\int_{S}\,dy\,dz$, where the density of the sink particles are excluded.
}\end{figure}

 \begin{figure}[t]
\includegraphics[scale=0.65]{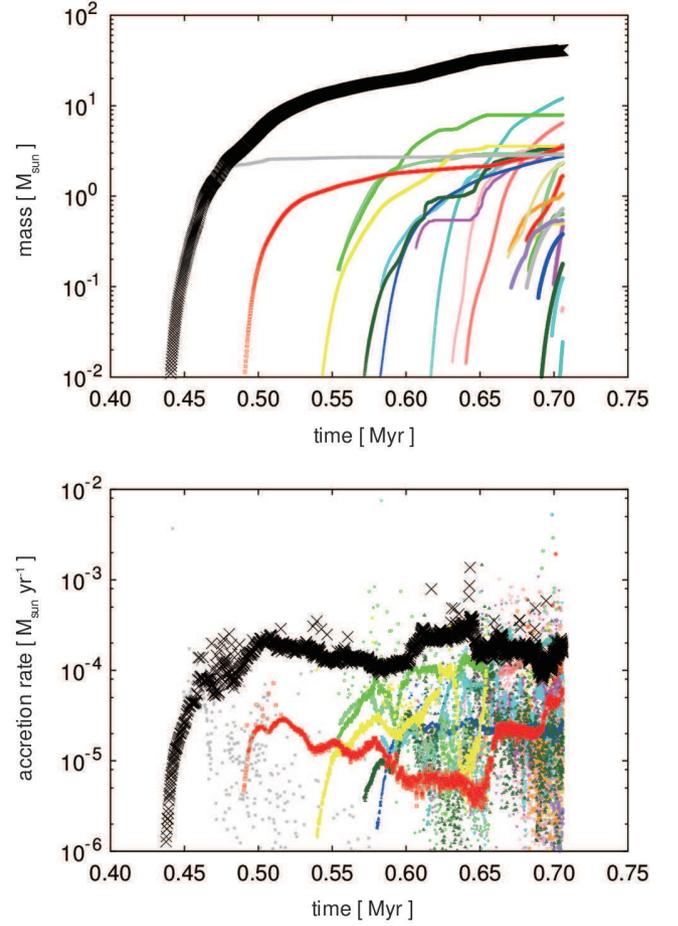}
\caption{\label{f7}
Evolution of mass ({\it top}) and mass accretion rate ({\it bottom}) of sink particles.
Black points corresponds to those of the most massive sink.
Other smaller mass sink particles are plotted with colors.
}\end{figure}

\begin{figure}[t]
\includegraphics[scale=0.60]{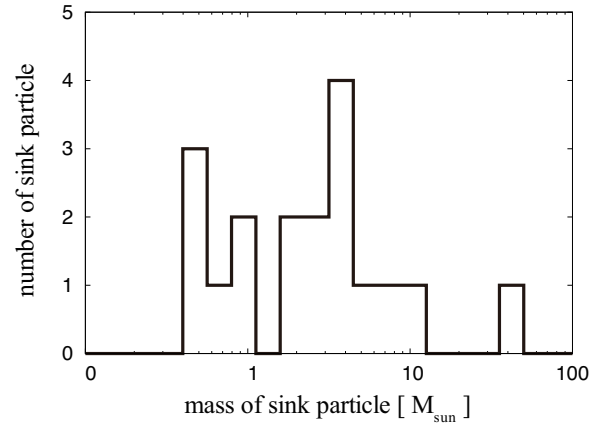}
\caption{\label{f9}
Mass distribution of all 19 sink particles at $t=0.70$ Myr
}\end{figure}

\begin{figure}[t]
\includegraphics[scale=0.65]{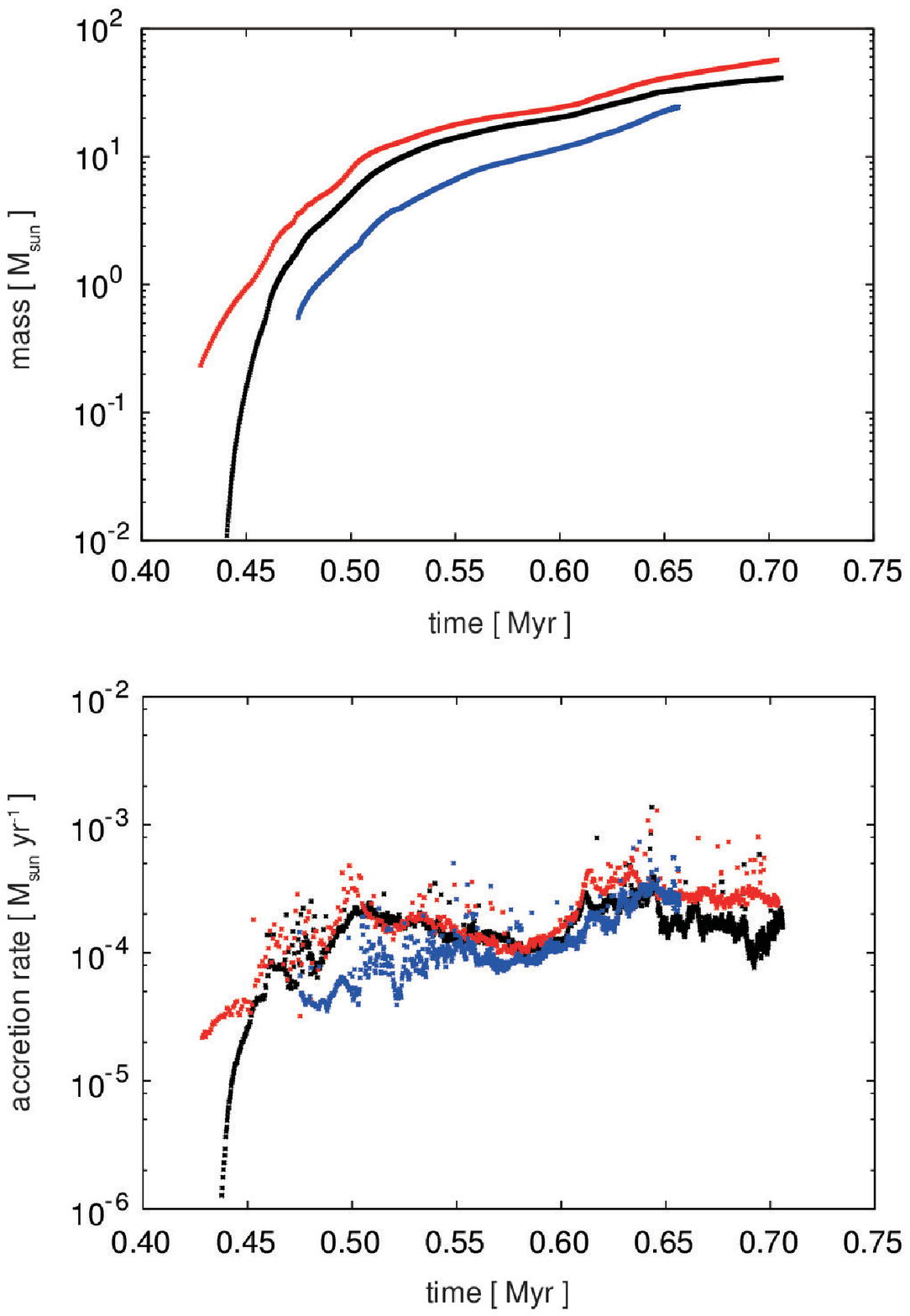}
\caption{\label{f8}
Evolution of mass ({\it top}) and mass accretion rate ({\it bottom}) of sink particles.
Black points corresponds to the most massive sink of the fiducial run.
Blue and red points show those of the higher and lower resolution runs, respectively.
}\end{figure}

In Figure \ref{f6}, to show that the most massive sink is fed by the global filament collapse, we show the velocity structure along the $x$-axis that passes by the sink position.
Top and bottom panels show the results of $t=0.45$ and $0.55$ Myr, respectively.
The purple lines show density-weighted average $v_x$ around the $x$-axis: $\langle v_x(x) \rangle_{y,z}\equiv\int_{S}\,\rho\,v_x\,dy\,dz/\int_{S}\,\rho\,dy\,dz$ where $S=\{(y,z)|\,[(y-y_{\rm sink})^2+(z-z_{\rm sink})^2]^{1/2}\le0.1\,\mbox{pc}\}$.
This velocity structure clearly shows that the filament is globally collapsing toward the sink.
The green and blue lines respectively indicate the velocity structures of $\langle v_x(x)+v_y(x) \rangle_{y,z}/\sqrt{2}$ and $\langle v_x(x)+v_z(x) \rangle_{y,z}/\sqrt{2}$.
If we observe this filament by using some line emissions from an oblique direction with the angle of 45$^\circ$ to the $z$-axis in the $x$-$y$ plane ($x$-$z$ plane), we would find the structure similar to the green (blue) line in position-velocity space.

Since the line-mass of the filament is supercritical, it would also be gravitationally unstable in smaller scales on the order of the filament width (Inutsuka \& Miyama 1992).
The oscillatory structures in the $\langle v_x(x) \rangle_{y,z}$ seem to be a signature of such gravitational instability.
In Figure \ref{f6} we plot the average number density structure of the filament: $\langle n(x) \rangle_{y,z}\equiv\int_{S}\,n\,dy\,dz/\int_{S}\,dy\,dz$ as the yellow lines, where the density of the sink particles are excluded.
We see that, in particular at $t=0.55$ Myr (bottom panel of Figure \ref{f6}), there are a few density peaks around the most massive sink that could be a consequence of the growth of smaller scale instabilities.
Note that, in the case of the unmagnetized filament, the growth rate of the instability is maximum at the scale about four times the diameter of the filament, and the growth rate decreases with increasing scale of the perturbations above the most unstable scale (Inutsuka \& Miyama 1992).
Under a strong magnetic field perpendicular to the filament like the present filament, the most unstable scale of the gravitational instability would become larger than the unmagnetized case because magnetic pressure enhances the effective sound speed in the direction along the filament.
A linear analysis of the filament with the perpendicular magnetic field is necessary to confirm this speculation.
Our result shows that the most massive sink particle experiences 5 coalescences with lower mass sink particles within a few times 0.1 Myr history of its evolution, indicating that both global collapse and local fragmentations occurred.

The evolution of mass and mass accretion rates of the sink particles are shown in the top and bottom panels of Figure \ref{f7}, respectively.
The black points correspond to those of the most massive sink.
The mass accretion rate keeps high value required to form O-type massive star ($\dot{M}\gtrsim10^{-4}\mbox{ M}_{\rm sun}$ yr$^{-1}$; Wolfire \& Cassinelli 1987) and the mass of the most massive sink exceeds $40\mbox{ M}_{\rm sun}$ within 0.3 Myr after its creation.
In Figure \ref{f9}, we show mass distribution of all 19 sink particles at $t=0.70$ Myr.

To test whether our massive sink formation process depends on the resolution, we examine additional runs with higher and lower resolutions.
In the higher resolution run, we change the Jeans criterion in the AMR condition from $f=8$ to $f=16$ (see, section 2) while keeping the maximum level of the refinement.
This modification leads twice better resolution for dense objects but sink particles can be created at lower densities than the fiducial model by a factor of 4.
In the lower resolution run, we reduce the maximum level of the AMR to be three, i.e., the maximum resolution is twice coarser than the fiducial run (this also leads to four time smaller threshold density for the sink formation).
The resulting masses of the most massive sink and the accretion rates for these runs are plotted in Figure \ref{f8}.
The figure indicates that at least the formation process of the most massive sink is not substantially affected by the resolution.

\section{Summary and Discussion}
We have studied the shock compression of a turbulent molecular clump using the AMR isothermal MHD simulations with self-gravity and sink particles.
We found that the compression leads to the formation of filaments and the longitudinal global collapse of the filament creates a massive sink particle with a mass larger than the 50 M$_{\rm sun}$.
This is particularly compatible with the cloud collision regions of S116 and NGC6334 (Fukui et al. 2017b, c) where the filamentary structures are suggested to be enhanced by the collision.
The mass accretion rate of the massive sink is lager than $10^{-4}$ M$_{\rm sun}$ yr$^{-1}$ indicating that even O-type star(s) can be formed within a million years after the shock passage.

The formation mechanism of the massive filament that bears the most massive sink is identified to be the focusing flows due to the curved MHD shock as proposed by Inoue \& Fukui (2013) (see also, Vaidya et al. 2013).
The line-mass of the filament is as large as $100$ M$_{\rm sun}$ pc$^{-1}$, which can be determined by the critical line-mass of the filament threaded by a strong magnetic field perpendicular to the filament with strength $\gtrsim 300\,\mu$G (see, eq.~[\ref{Mfil}]).
This is compatible with the recent observation of a massive filamentary infrared dark cloud G11.11$-$0.12 (Pillai et al. 2015).
 
Is the above massive star formation mechanism applicable only for a cloud collision scenario?
We believe that we can naturally expect similar massive star formation modes especially in giant molecular clouds.
This is because our mechanism can work if there is a shock wave that is strong enough to create a massive filament.
According to Larson's law, which suggests higher velocity dispersion for larger cloud, high Mach number shock would be ubiquitous in giant molecular clouds due to its own turbulence.
Our future study will examine this expectation.
 
In this paper, we have studied the head-on collision between a small and a large cloud.
Is this a realistic situation?
Observations and theories suggest that the mass function of molecular clouds and the mass function of clumps in molecular clouds show power-law shape $dN/dm\propto m^{-a}$ with the spectral index roughly $a\sim 1.7$ (Kramer et al. 1996; Fukui et al. 2008; Hennebelle \& Chabrier~2008; Inoue \& Inutsuka~2012; Inutsuka et al.~2015; Kobayashi et al.~2017).
Hence the rate of collision between a molecular cloud with mass $M$ and a cloud with mass larger than $M$ can be estimated as 
\begin{eqnarray}
R(M)=\int_M\,\sigma\,v_{\rm coll}\,m^{-a}\,dm \propto M^{-a+q+1},
\end{eqnarray}
where $\sigma\propto m^{q}$ is the cross section of the collision (spherical cloud has q=2/3 and constant surface density cloud corresponds to q=1) and we have assumed that the collision velocity does not depend on the mass, for simplicity.
Both cases of $q$ show very weak dependence of $R(M)$ on $M$, implying that the collision rate does not substantially depend on the mass $M$.
Thus the collision of the clouds with different mass scales and with similar mass scales happen with a similar rate.
When we consider the collision of similar mass clouds, we need to pay attention to offset collisions that could induce a shear flow in the shocked layer and that potentially change the dynamics of the filament formation.

In this paper, we created the sink particle when the local density reaches about $3\times10^{-17}$ g cm$^{-3}$, which is much smaller than that of the first adiabatic core $\sim 10^{-13}$ g cm$^{-3}$.
Gravitational instability may occur in the first core and possibly lead to the formation of multiple stars. Thus, our sink mass should be interpreted as the upper limit of the mass of a massive star.
We clearly need future studies with higher resolution and with radiation feedback to conclude rigorous massive star formation.

T. I. acknowledges D. Arzoumanian, T. Hosokawa, and J. Tan for fruitful scientific conversations.
The numerical computations were carried out on XC30 system at the Center for Computational Astrophysics (CfCA) of National Astronomical Observatory of Japan.
This work is supported by Grant-in-aids from the Ministry of Education, Culture, Sports, Science, and Technology (MEXT) of Japan, No.~15K05039 (T.I.), and No.~17K05394 and 17H02863 (T.M.), and supported by NAOJ abroad visiting program for young researcher (T.I. and P.H.).


\begin{thebibliography}{}
\bibitem[Anathpindika(2010)]{1} Anathpindika, S. V. 2010, \mnras, 405, 1431
\bibitem[Arzoumanian et al.(2011)]{3} Arzoumanian, D., et al. 2011, \aap, 529, L6
\bibitem[Balfour et al.(2017)]{BWH17} Balfour, S. K., Whitworth, A. P., \& Hubber, D. A. 2017, \mnras, 465, 3483
\bibitem[Bonnel et al.(2001)]{5} Bonnell, I. A., Bate, M. R., Clarke, C. J., \& Pringle, J. E. 2001, \mnras, 323, 785
\bibitem[Chen \& Ostriker(2014)]{CO14} Chen, C.-Y. \& Ostriker, E. C. 2014, \apj, 785, 69
\bibitem[Chen \& Ostriker(2015)]{CO15} Chen, C.-Y. \& Ostriker, E. C. 2015, \apj, 810, 126
\bibitem[Commercon et al.(2011)]{7} Commercon, B., Hennebelle, P., \& Henning, T. 2011, \aap, 530, 13
\bibitem[Crutcher et al.(2010)]{C10} Crutcher, R. M., Wandelt, B., Heiles, C., Falgarone, E., \& Troland, T. H. 2010, \apj, 725, 466
\bibitem[Dedner et al.(2002)]{D02} Dedner, A., Kemm, F., Kroner, D., et al. 2002, J. Comp. Phys., 175, 645
\bibitem[Dobashi et al.(2014)]{D14} Dobashi, K., Matsumoto, T., \& Shimoikura, T. et al. 2014, \apj, 797, 58
\bibitem[[Federrath et al.(2010)]{F10} Federrath, C., Banerjee, R., Clark, P. C., \& Klessen, R. S. 2010, \apj, 713, 269
\bibitem[Fukui et al.(2008)]{F08} Fukui, Y., Kawamura, A., Minamidani, T. et al. 2008, \apjs, 178, 56
\bibitem[Fukui et al.(2014)]{F14} Fukui, Y., Ohama, A., Hanaoka, N. et al. 2014, \apj, 780, 36
\bibitem[Fukui et al.(2015)]{F15} Fukui, Y. et al. 2015, \apj, 807, 4
\bibitem[Fukui et al.(2016)]{F16} Fukui, Y. et al. 2016, \apj, 820, 26
\bibitem[Fukui et al.(2017a)]{F17a} Fukui, Y. et al. 2017a, arXiv:1706.05768
\bibitem[Fukui et al.(2017b)]{F17b} Fukui, Y. et al. 2017b, arXiv:1706.05771
\bibitem[Fukui et al.(2017c)]{F17c} Fukui, Y. et al. 2017c, arXiv:1706.08720
\bibitem[Furukawa et al.(2009)]{10} Furukawa, N., Dawson, J. R., Ohama, A. et al. 2009, \apj, 696, 115
\bibitem[Galv\'{a}n-Madrid et al.(2010)]{G10} Galv\'{a}n-Madrid, R., Zhang, Q., Keto, E. et al. 2010, \apj, 725, 17
\bibitem[Girichidis et al.(2011)]{G11} Girichidis, P., Federrath, C., Banerjee, R., \& Klessen, R. S. 2011, \mnras, 413, 2741
\bibitem[Habe \& Ohta(1992)]{11} Habe, A. \& Ohta, K. 1992, \pasj, 44, 203
\bibitem[Hasegawa et al.(1994)]{H94} Hasegawa, T., Sato, F., Whiteoak, J. B., \& Miyawaki, R. 1994, \apj, 429, 77
\bibitem[Hayashi et al.(2017)]{H17} Hayashi, K. et al. 2017a, arXiv:1706.05871
\bibitem[Hennebelle \& Chabrier(2008)]{HC08}  Hennebelle, P., \& Chabrier, G. 2008, \apj, 684, 395
\bibitem[Hennebelle et al.(2011)]{39}  Hennebelle, P., Commercon, B., Joos, M. et al. 2011, \aap, 528, 72
\bibitem[Higuchi et al.(2014)]{H14} Higuchi, A. E., Chibueze, J. O., Habe, A., Takahira, K., \& Takano, S. 2014, \apj, 147, 141
\bibitem[Inoue \& Inutsuka (2012)]{14} Inoue, T., \& Inutsuka, S. 2012, \apj, 759, 35
\bibitem[Inoue \& Fukui (2013)]{IF13} Inoue, T., \& Fukui, Y. 2013, \apj, 774, 31
\bibitem[Inutsuka \& Miyama (1992)]{IM92} Inutsuka, S. \& Miyama, S. M. 1992, \apj, 388, 392
\bibitem[Inutsuka et al.(2015)]{II15} Inutsuka, S., Inoue, T., Iwasaki, K., \& Hosokawa, T. 2015, \apj, 580, 49
\bibitem[Ohama et al.(2017a)]{O17} Ohama, A. et al. 2017a, arXiv:1706.05652
\bibitem[Ohama et al.(2017b)]{O17} Ohama, A. et al. 2017b, arXiv:1706.05659
\bibitem[Ostriker 1964]{O64} Ostriker, J. P. 1964, \apj, 140, 1056
\bibitem[Takahira et al.(2014)]{TTH14} Takahira, K., Tasker, E. J., \& Habe, A. 2014, \apj, 792, 63
\bibitem[Tan et al.(2014)]{Tppvi14} Tan, J. C. et al. 2014, , in Protostars and Planets VI (Tuscon, AZ: Univ. Arizona Press), 149
\bibitem[Tomisaka (2014)]{T14} Tomisaka, K. 2014, \apj, 785, 24
\bibitem[Tsutsumi et al.(2017)]{O17} Tsutsumi, D. et al. 2017, arXiv:1706.05664
\bibitem[Kohno et al.(2017)]{K17} Kohno, M. et al. 2017, arXiv:1706.07964
\bibitem[Kobayashi et al.(2017)]{KIKH17} Kobayashi, M. I. N., Inutsuka, S., Kobayashi, H., \& Hasegawa, K. 2017, \apj, 836, 175
\bibitem[Kramer et al.(1996)]{K96} Kramer, C., Stutzki, J., \& Winnewisser, G. 1996, \aap, 307, 915
\bibitem[Krumholz et al.(2009)]{18} Krumholz, M. R., Klein, R. I., McKee, C. F., Offner, S. S. R., \& Cunningham, A. J. 2009, Science, 323, 754
\bibitem[Kuiper et al.(2010)]{KKBH10} Kuiper, R., Klahr, H., Beuther, H., \& Henning, T. 2010, \apj, 722, 1556
\bibitem[Larson(1981)]{32} Larson, R. B. 1981, \mnras, 194, 809
\bibitem[Matsumoto (2007)]{M07} Matsumoto, T. 2007, \pasj, 59, 905
\bibitem[Matsumoto et al.(2015)]{MDS15} Matsumoto, T., Dobashi, K., \& Shimoikura, T. 2015, \apj, 801, 77
\bibitem[Miyoshi \& Kusano(2005)]{MK05} Miyoshi, T. \& Kusano, K. 2005, J. Com. Phys., 208, 315
\bibitem[Mouschovias \& Spitzer(1976)]{MS76} Mouschovias, T. Ch., \& Spitzer, L., Jr. 1976, \apj, 210, 326
\bibitem[Myers et al.(2013)]{21} Myers, A. T., McKee, C. F., Cunningham, A. J., Klein, R. I., \& Krumholz, M. R. 2013, \apj, 766, 97
\bibitem[Nakano et al.(2000)]{22} Nakano, T., Hasegawa, T., Morino, J., \& Yamashita, T. 2000, \apj, 534, 976
\bibitem[Nishimura et al.(2017a)]{N17a} Nishimura, A. et al. 2017a, arXiv:1706.06002
\bibitem[Nishimura et al.(2017b)]{N17b} Nishimura, A. et al. 2017b, arXiv:1706.06956
\bibitem[Ohama et al.(2010)]{23} Ohama, A., Dawson, J. R., Furukawa, N. et al. 2010, \apj, 709, 975
\bibitem[Peretto et al.(2007)]{P07} Peretto, N., Hennebelle, P., \& Andr\'e, P. 2007, \aap, 464, 983
\bibitem[Peretto et al.(2013)]{P13} Peretto, N., Fuller, G. A., Duarte-Cabral, A. et al. 2013, \aap, 555, 112
\bibitem[Peretto et al.(2014)]{P14} Peretto, N., Fuller, G. A., Andr\'e, P. 2014, \aap, 561, 83
\bibitem[Peters et al.(2011)]{38} Peters, T., Banerjee, R., Klessen, R. S., \& Mac Low, M.-M. 2011, \apj, 729, 72
\bibitem[Pillai et al.(2015)]{PKT15} Pillai, T., Kauffmann, J., Tan, J. C., Goldsmith, P. F., Carey, S. J., \& Menten, K. M. 2015, \apj, 799, 74
\bibitem[Sano et al.(2017)]{O17} Sano, H. et al. 2017, arXiv:1706.05763
\bibitem[Shima et al.(2017)]{STH17} Shima, K., Tasker, E. J., \& Habe, A. 2017, arXiv:1612.06381
\bibitem[Stodolkiewicz 1963]{S63} Stodolkiewicz, J. S. 1963, AcA, 13, 30
\bibitem[Torii et al.(2011)]{25} Torii, K., Enokiya, R., Sano, H. et al. 2011, \apj, 738, 46
\bibitem[Torii et al.(2015)]{T15} Torii, K., Hasegawa, K., Hattori, Y. et al. 2015, \apj, 806, 7
\bibitem[Torii et al.(2017)]{T17} Torii, K. et al. 2017, arXiv:1706.07164
\bibitem[Trulove et al.(1997)]{26} Truelove, J. K., Klein, R. I., McKee, C. F. et al. 1997, \apj, 489, 179
\bibitem[Tsuboi et al.(2015)]{TMU15} Tsuboi, M., Miyazaki, A., \& Uehara, K. 2015, \apj, 67, 109
\bibitem[Vaidya et al.(2013)]{V13} Vaidya, B., Hartquist, T. W., \& Falle, S. A. E. G. 2013, \mnras, 433, 1258
\bibitem[Wolfire \& Cassinelli(1987)]{29} Wolfire, M. G., \& Cassinelli, J. P. 1987, \apj, 319, 850
\bibitem[Wu et al.(2017)]{WB17} Wu, B., et al. 2017a, \apj, 835, 137
\bibitem[Wu et al.(2017)]{WB17} Wu, B., et al. 2017b, arXiv:1702.08117
\bibitem[Yorke \& Sonnhalter(2002)]{30} Yorke, H. W., \& Sonnhalter, C. 2002, \apj, 569, 846
\bibitem[Zinnecker \& Yorke(2007)]{31} Zinnecker, H., Yorke, H. W. 2007, \araa, 45, 481
\end{thebibliography}
\end{document}